\documentclass[multphys,vecphys]{svmult}

\usepackage{makeidx}         
\usepackage{graphicx}        
\usepackage{multicol}        
\usepackage[bottom]{footmisc}

\makeindex             

\begin{document}

\title*{First AMBER/VLTI observations of hot massive stars}

\titlerunning{Hot massive stars with AMBER on the VLTI}
\author{R.\,G.\,Petrov \inst{1}\and F.\,Millour \inst{1,2} \and O.\,Chesneau \inst{3} \and G.\,Weigelt \inst{4}\and D.\,Bonneau \inst{3} \and Ph.\,Stee \inst{3} \and S.\,Kraus \inst{4}\and D.\,Mourard \inst{3} \and A.\,Meilland \inst{3} \and M.\,Vannier \inst{5}\and F.\,Malbet \inst{2} \and F.\,Lisi \inst{6} \and P.\,Antonelli \inst{3}and P.\,Kern \inst{2} \and U.\,Beckmann \inst{4} \and S.\,Lagarde \inst{3}\and K.\,Perraut \inst{2} \and S.\,Gennari \inst{5} \and E.\,Lecoarer \inst{2}\and Th.\,Driebe \inst{4} \and M.\,Accardo \inst{5} \and S.\,Robbe-Dubois \inst{1}\and K.\,Ohnaka \inst{4} \and S.\,Busoni \inst{6} \and A.\,Roussel \inst{3}\and G.\, Zins\inst{2} \and J.\, Behrend\inst{4} \and D.\, Ferruzi\inst{5} \and Y.\, Bresson\inst{3} \and G.\, Duvert\inst{2} \and E.\, Nussbaum\inst{4} \and A.\, Marconi\inst{5} \and Ph.\, Feautrier\inst{2} \and M.\, Dugu\'{e}\inst{3} \and A.\, Chelli\inst{2} \and E.\, Tatulli\inst{2} \and M.\, Heininger\inst{4} \and A.\, Delboulbe\inst{2} \and S.\, Bonhomme\inst{3} \and D.\, Schertl\inst{4} \and L.\, Testi\inst{6} \and Ph.\, Mathias\inst{3} \and J.\,-L.\, Monin\inst{2} \and L.\, Gluck\inst{2} \and K.\,H.\, Hofmann\inst{4} \and P.\, Salinari\inst{6} \and P.\,Puget\inst{2} \and J.\,-M.\, Clausse\inst{3} \and D.\, Fraix-Burnet\inst{2} \and R.\, Foy\inst{7} \and A.\, Isella\inst{6}}
\authorrunning{R.G. Petrov and the AMBER Consortium}
\institute{LUAN, Universit\'{e} de Nice - Sophia Antipolis. Parc Valrose, F-06108 Nice Cedex 2, France.
\texttt{petrov@unice.fr}
\and {LAOG, Universit\'{e} Joseph Fourier, BP 53, F-38041 Grenoble Cedex 9, France}
\and{Observatoire de la C\^{o}te d'Azur,BP
4229, F-06304 Nice Cedex 4, France}
\and{Max Planck Institute f\"{u}r Radioastronomie, D-53121 Bonn, Germany}
\and{European Southern Observatory, Casilla 19001, Santiago, Chile}
\and{Osservatorio Astrofisico di Arcetri, Istituto Nazionale di Astrofisica, Largo E.
Fermi 5, I-50125 Firenze, Italy}\and{CRAL, Observatoire de Lyon, F-69561 Saint Genis Laval Cedex, France}
}

%
\maketitle
\abstract AMBER is the first near infrared focal instrument of the VLTI. It combines three telescopes and produces spectrally resolved interferometric measures. This paper discusses some preliminary results of the first scientific observations of AMBER with three Unit Telescopes at medium (1500) and high (12000) spectral resolution. We derive a first set of constraints on the structure of the circumstellar material around the Wolf Rayet $\gamma^{2}$ Velorum and the LBV $\eta$ Carinae.
\vskip -0.15cm
\section{Introduction}
\label{sec:1}
\vskip -0.2cm
A feature common to many hot and massive stars is a complex circumstellar envelope revealed by strong emission lines in the spectrum and excesses in the continuum Spectral Energy Distribution (SED). The classical Be star $\kappa$ Canis Majoris, the brightest Wolf Rayet  $\gamma^{2}$ Velorum and the more than famous Luminous Blue Variable $\eta$ Carinae were among the first medium spectral resolution AMBER Guaranteed Time Targets.\

These stars belong to fairly different classes, with, to make it short, specific evolution stages, extremely different mass loss rates (respectively about $10^{-9}$, $10^{-5}$ and $10^{-3} M_{\odot}/y$), different envelope densities, opacities and chemical compositions.\

However, they raise variants of common questions. What are the exact mass loss rates, since their computation requires a model of the envelope geometry? Can we discriminate from the contribution of dust and of free-free emission of gas to the continuum spectrum? Are they mechanisms allowing dust to be present closer to the star than the expected sublimation radius? What are the relative contributions of radiation pressure and stellar rotation to the production and shaping of the envelope? Can we confirm that the radiation pressures show very strong variations with latitude, in particular with the recently renewed importance of the Von Zeipel effect increasing the apparent gravitation and radiation pressure near the poles of fast rotators? Are the stars close enough to critical velocity for this to be the main explanation for mass loss and even for variability through a rotational instability mechanism? Do we have to look into stellar activity producing local perturbations of the velocity field and/or of the radiation pressure? Are there other eruption mechanisms? Is the envelope completely shaped by the emission mechanisms (wind, rotation, kinetic momentum transfer) or is it severely perturbed by other sources such as binarity, which is of course decisive for $\gamma^{2}$ Velorum but also suspected to be important for $\eta$ Carinae. In the influence of the companion on the circumstellar material what is the part of gravitation and this of the companions own stellar wind and can we see a wind-wind interaction zone? It is now quite clear that spherically symmetric models are outdated but can we still hold on central symmetric ones, eventually moderately perturbed?\

It seems that a full answer to these questions would require full images of the targets in the continuum as well as in many narrow spectral channels in many different emissions lines, allowing to derive intensity maps and velocity fields at different optical depths. In principle, AMBER is able to produce such color images in the near infrared thanks to its three telescopes beam combiner feeding a medium (R=1500) and high (R=12000) resolution spectrograph. They are the goal of long term programs of AMBER which will require a very large number of observations with the Auxiliary 1.8 meter telescopes. In the meantime, we decided that these objects are an excellent test case for one of the main bets behind the conception of the "only three telescopes but ambitious spectrograph" AMBER instrument. Since such most studied candidates are already fairly constrained by multi wavelength (from X to far IR) spectro-photometry, high resolution spectroscopy, larger scale imaging, polarimetry and often interferometry without spectral resolution, then a small number of interferometric measures simultaneously in a large number of spectral channels, should allow decisive breakthroughs. What we have measured is very little information compared to full color images but it also multiplies purely spectroscopic information by at least a factor 8.\

In the following, we give a first insight of our very preliminary understanding of the first AMBER measures made on these stars.\
\section{AMBER observations}
\vskip -0.2cm
AMBER is a three beams near infrared VLTI focal instrument producing dispersed fringes with spectral resolutions 35, 1500 and 12000 \cite{petrov03}. This paper refers to medium and high spectral resolution observations made in the K band. Figure 1 displays an example of the individual image detected by the AMBER detector, with left to right the photometric 1 and 2 channels, then the interferometric one and the photometry of the third beam, all dispersed in the vertical direction. One can see a three telescopes fringe pattern in the interferometric channel and the $Br_{\gamma}$ emission line crossing all spectra horizontally. The figure also describes the work channel and the reference channel.
AMBER measures the stellar spectra, the absolute visibility in each channel, the differential visibility which makes sense even when the absolute visibility is poorly calibrated, the differential phase and the closure phase. These quantities can be interpreted in an unique manner only if we have a good u-v coverage. However, it can be remembered that the visibility is related to the angular scale in  $\lambda/B_i$ units in the direction of the baseline $B_i$. The differential phase gives an idea of the position of the object photocenter in the direction of the baseline $B_i$ in the spectral channel $\lambda$ . This is particularly true when the phase is smaller than 1 radian. The closure phase is a measure of the asymmetry of the source: central symmetric or unresolved sources have a zero closure phase. These generalities can be wrong in an infinity of particular cases in which they are wrong, but they still are quite useful for a first interpretation and initial orientation of the model fitting.
\begin{figure}[h]
\centering
\includegraphics[width=11.5cm]{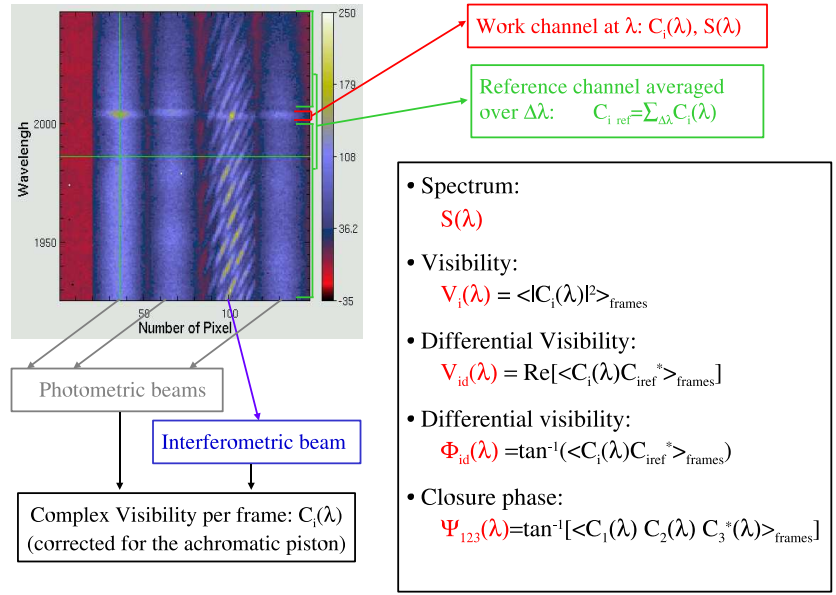}

\caption{\scriptsize{
AMBER typical bright star, three telescopes, medium spectral resolution image and AMBER measures.}}
\label{petrov:fig:2}
\end{figure}

For bright sources we currently can guarantee an absolute visibility accuracy ranging between 0.03 and 0.06 and differential visibility and phase smaller than 0.01 (visibility units and radians respectively). The closure phase is usually substantially more noisy than the added individual phases, because the probability to have simultaneously three good fringe patterns is quite low in the currently vibration dominated VLTI.
\section{The Wolf Rayet star $\gamma^{2}$ Velorum}
\vskip -0.2cm
We observed the brightest Wolf Rayet $\gamma^{2}$ Velorum in a set of spectral windows in the K band from 1.95 to 2.17 $\mu m$. This star have been very extensively observed by spectroscopy, polarimetry and intensity interferometry \cite{hanbury} and is known to be a SB2 spectroscopic binary system (WC8+O7.5III) with period 78.53 days, $(a1+a2)\sin i=164.10^{6}$ Km, eccentricity =0.326. The inclination i=$63\pm5 deg$ is fairly tightly constrained by intensity interferometry and polarimetry \cite{schmutz}. The Hipparcos distance is $258\pm40$ pc and there is a not completely closed controversy with pre-Hipparcos spectro-photometric estimations which were around 400 pc.
After constraining again the orbit and the distance by a new interferometric measurement the main goal was to find out where is the circumstellar material in the system. The individual stars should be too small to be resolved (less than 0.5 mas) even if the W.R. seems larger because of the optically thick wind. However, we can expect some amount of dust in the system even if SED fits indicate that this should not yield more than 10\% of the total flux. We also expect some gas concentrated in the zone where the radiation pressure of the two stars are comparable. This wind-wind zone can contribute to the emission lines but also to the continuum through a free-free emission. The results are displayed if figure 2 with, clockwise from the upper left corner: the spectrum, showing many carbon and helium emission lines, the closure phase, the differential phases and the visibilities as a function of wavelength. The main spectral features are indicated in the spectrum, which is dominated by two strong emission groups at 2.079  m ($C IV$) and 2.11 ($He I$) both strongly blended. The closure phase is of the order of 1 radian in the "continuum" channels and shows very strong variations in the emission lines. So does the differential phases, which are set to a zero average in each observing window by the definition of the reference channel, but which very strongly vary in the lines.
\begin{figure}[h]
\centering
\includegraphics[angle=0, width=5.8cm]{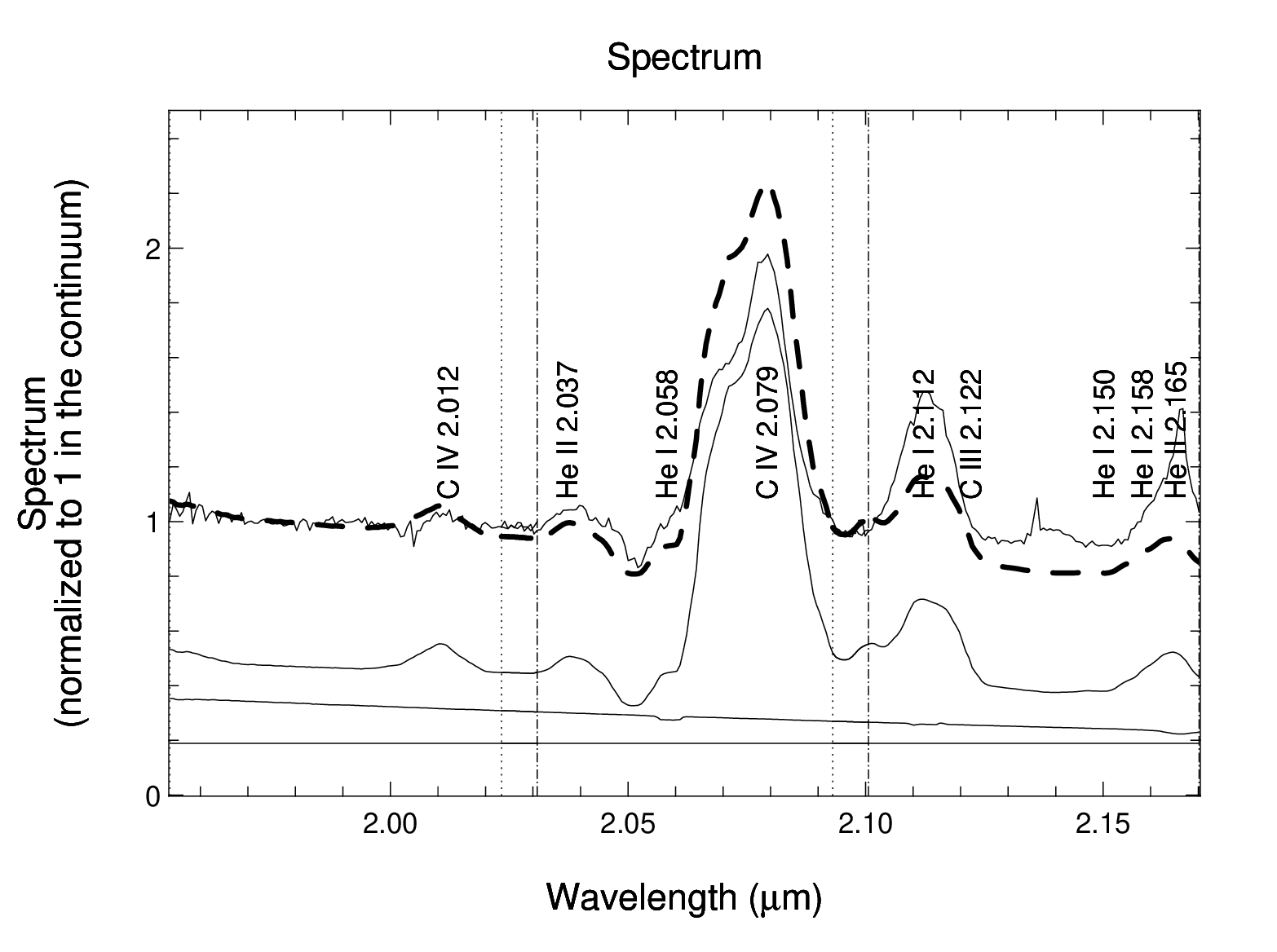}
\includegraphics[angle=0, width=5.8cm]{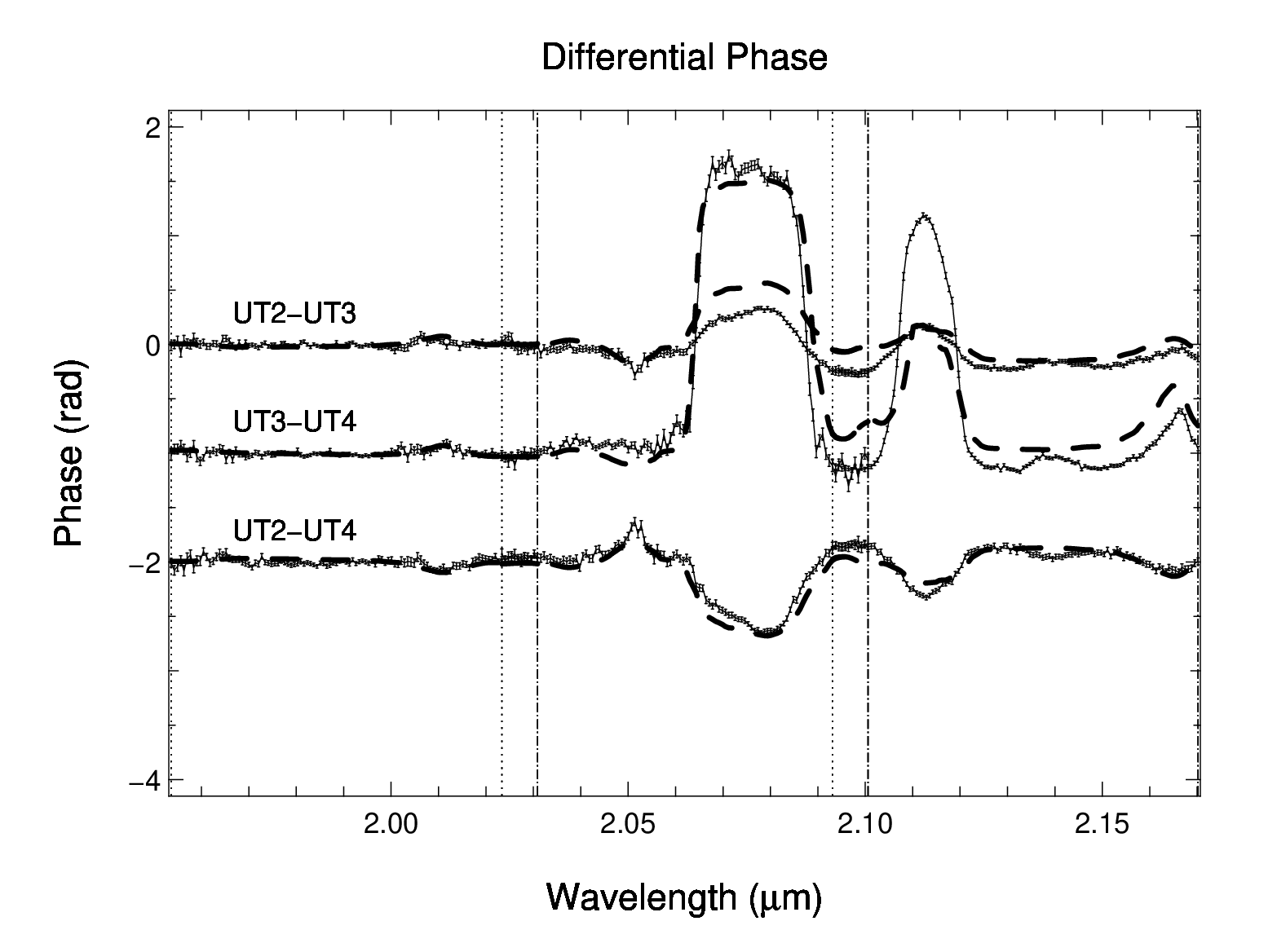}
\includegraphics[angle=0, width=5.8cm]{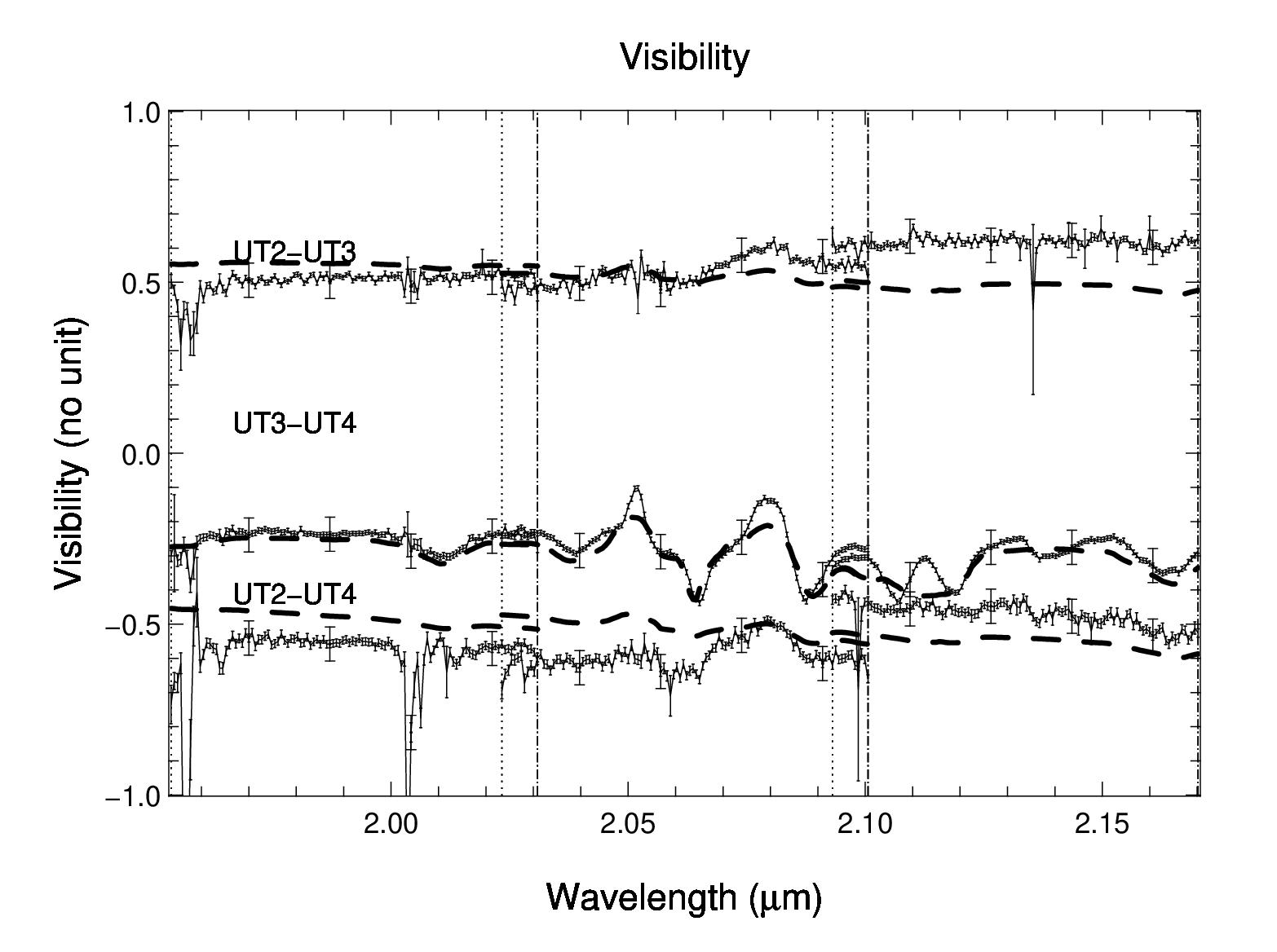}
\includegraphics[angle=0, width=5.8cm]{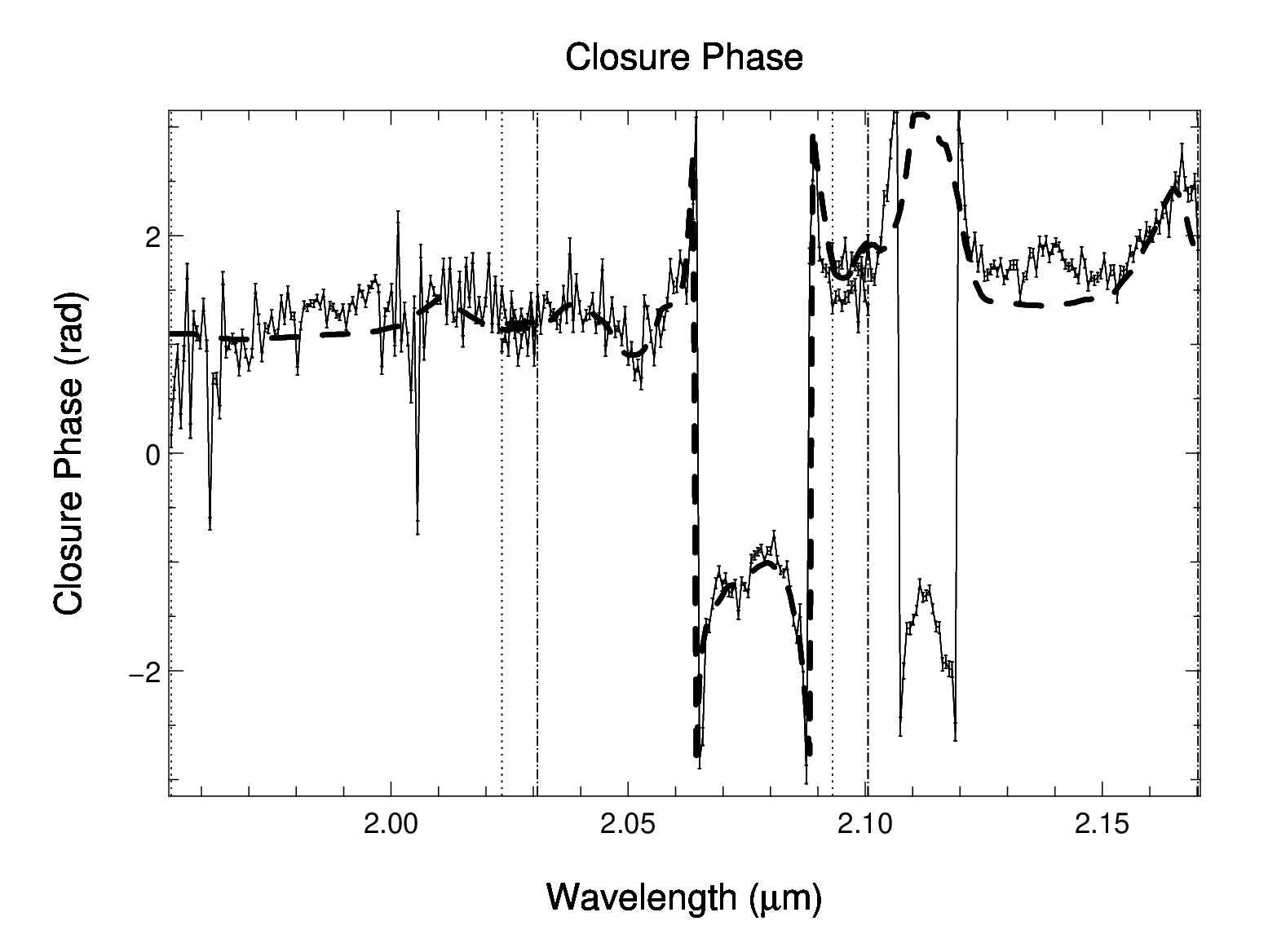}
\caption{\scriptsize{
AMBER observations of $\gamma^{2} Velorum$. The vertical lines show the limit of the observing spectral windows. The spectrum displays from top to bottom: the calibrated normalized spectrum (full line), the best fit (dashed line), the WC8 component model spectrum, the O7III component model spectrum and the Resolved Component flat contribution. In the visibility figure we have V (top) for UT2-UT3, V-0.5 (middle) for UT3-UT4 and V-1 (bottom) for UT2-UT4. The small error bars are for differential visibility accuracy and the large ones are from the external calibration uncertainty. For the differential phases we have $\Delta\phi$ for UT2-UT3,  $\Delta\phi-1$    for UT3-UT4 and   $\Delta\phi-2$   for UT3-UT4. In all cases, the full line represents measures or inputs to the model and the dashed line is for the best fit with a binary.}}
\label{petrov:fig:2}
\end{figure}
To fit the measurements, we are performing the following steps:
\begin{itemize}
\item {Try to find the binary parameters in "continuum" channels. This failed even with increased error bars. A solution is to introduce a faint Resolved Component (RC), contributing to 15\% of the total flux Then we have a very good fit.}
\item {Use models of the component spectra, derived from spectroscopic observations \cite{dessart}, to further constrain the binary parameters. This is what is represented in figure 2 and yields the parameter which are discussed below. Remarkably, we find similar binary and RC parameters in the continuum and in the lines.}
\item {Improve the fit by changing the WR spectrum. Actually we will derive the best possible WR spectrum knowing the parameters of the binary. This step is currently under progress and improves the fit of the closure and differential phases. From the observation of figure 2, it already appears that the phases are well fitted where the WR spectrum reconstruction is coherent, for example around the $C IV$ 2.079 $\mu$m line, and wrong where the intensity of the WR line is incorrect as it can be seen in the spectrum near 2.112 $\mu$m. However, it is also clear that a change of the component spectra cannot explain the sharp variation in visibility which can be seen for example for the UT3-UT4 baseline at 2.053, 2.081, 2.108 and 2.12 $\mu$m,. The strong visibility variations reveal small scale structures, possibly with large radial velocities since they seem to be in the wings of the C IV and He I lines. We believe that this is the signature of circumstellar material concentrated near the wind-wind collision zone.}
\item {Constrain the geometry and kinematics of the wind-wind zone by fitting the visibility, closure and differential phase residuals left after the best binary fit. These residuals are 3 to 5 times larger than the differential error bars. First, we will try to introduce in each spectral channel a Gaussian "cloud" whose size, intensity and position might materialize the corresponding isovelocity wind-wind region. Second, we will use the supposedly known geometry of the wind region around the axis joining the two stars to have a fit with less parameters.}
\item {From the geometry of the wind region, we can derive its contribution to the continuum spectrum. Then, we will discriminate dust and gas contributions to the continuum and repeat all the process.}
\item {Eventually, all this relatively simple mix of geometrical constraints on physical parameters might yield a global model of the system, whose parameters will again be evaluated from the measures.}
\end{itemize}
\begin{table}
\centering
\caption{\scriptsize{Parameters of the $\gamma^{2}$ Vel binary system}}
\label{tab:1}       
%
%
\begin{tabular}{llll}
\hline\noalign{\smallskip}
separation $\rho$      . & position angle      . & flux ratio      . & resolved component    .\\
\noalign{\smallskip}\hline\noalign{\smallskip}
$3.6\pm0.03  mas   $ & $165\pm5 \deg   $ & $0.6\pm0.1$ & $0.14\pm0.04$ \\
\noalign{\smallskip}\hline
\end{tabular}
\end{table}
The best global fit is obtained for the binary and resolved (RC) component parameters given in table 1. This has to be compared with the  $\rho=5 mas$ and  $\theta=161 deg$ expected from the spectroscopic binary and the Hipparcos distance. The new value of the separation can be explained by an error on the inclination $\sin i$ and/or on the parallax. Since $\sin i$ seems much better constrained, our results yields a parallax of $348\pm50 pc$ which is compatible both with the Hipparcos estimate (we are at 2.2 $\sigma$) and with the pre-Hipparcos spectro-photometry.
\section{The Luminous Blue Variable $\eta$ Carinae }
\vskip -0.2cm
The Luminous Blue Variable $\eta$ Carinae, most luminous star known in our Galaxy, is one of the best studied and maybe less understood massive stars. It is the source of the spectacular Homunculus nebulae, produced by a maybe 10 solar masses outburst in 1840. A second massive outburst occurred in 1890 and produced a second nebulae which seems to be an embedded smaller replica of the first one. The reason for the outburst remains unknown but many attempts are being made to connect the shapes of the present day stellar wind and the nebulae general geometry. After indications from the HST Imaging Spectrograph, the first VLTI observations with VINCI strongly indicated that the stellar wind appears elongated in the nebulae axial direction which is also believed to be the stellar rotation axis. This is compatible with a radiation pressure substantially increased near the poles by a Von Zeipel effect. Our goal was to extend the VINCI observations by adding spectral resolution to the interferometric measures in order to constrain the velocity field.

One of the main difficulties in observing $\eta$ Car with single mode fiber instruments such as AMBER or VINCI is that the fibers collect information from an extended patch of the sky. In the case of VINCI siderostats, an array of about 1.4 arc second contribute to the interferogram. Images from the NACO adaptive optics (resolution 50 to 100 mas) have been used to find out what fraction of the collected flux can actually contribute to the fringes (i.e. is produced in an array smaller than 10 to 50 mas). This allowed to estimate that the central source contains 57\% of the flux. When it was assumed that the remaining 43\% are completely resolved by the interferometer, the resulting visibilities show a smooth variation with the position angle of the baseline. This is interpreted as a present day wind shape elongated exactly along the axis of the nebulae, with a ratio of 1.25 between the projected major and minor axes.

The first result of the AMBER 3 UT measurements is that closure phase is zero (within the 0.05 radians accuracy) in the continuum for all observing times, covering a fairly large range of hour angles. This is a good confirmation that at least in the continuum the object is well represented by a central symmetric structure such as the Gaussian ellipsoid.

The UTs inject in AMBER fibers the light coming from about 70 mas. We first assumed that AMBER signal would be completely dominated by the central stellar wind. Figure 3a shows the AMBER measures (triangles), each visibility point converted in the FWHM of a Gaussian. The dashed line shows the Van Boekel et al. Vinci fit \cite{vanboekel}. The AMBER points are not compatible with the VINCI fit and show strong variations of Gaussian FWHM with small variations of position angle. If we assume that AMBER data has been contaminated by a fraction of light for an interferometrically resolved  source, then it is possible to eliminate the strong variations of FWHM with PA. The remarkable point is that we then obtain a structure elongated exactly in the same direction and with the almost the same major to minor axis ratio than from VINCI. Figure 3b shows the AMBER measures corrected assuming that the flux ratio between the resolved and central structures is 0.45 (i.e. the resolved structure contributes to 31\% of the total flux instead of 43\% in the VINCI case). The dotted curve shows the VINCI fit, scaled down by a factor 1.33. The easiest way to explain this difference in apparent size is to challenge the VINCI estimation of the contribution of non resolved structures to the total flux, since we now know that a fraction of the flux in one UT Airy disk comes from structures non resolved by the interferometer . The values in figure 3b are indeed compatible with VINCI measurement where 47\% instead of 43\% of the flux contribute to the unresolved structure. This would slightly change the major to minor axis ratio but not the position angle of the structure.

\begin{figure}[h]
\centering
\includegraphics[width=5.7cm]{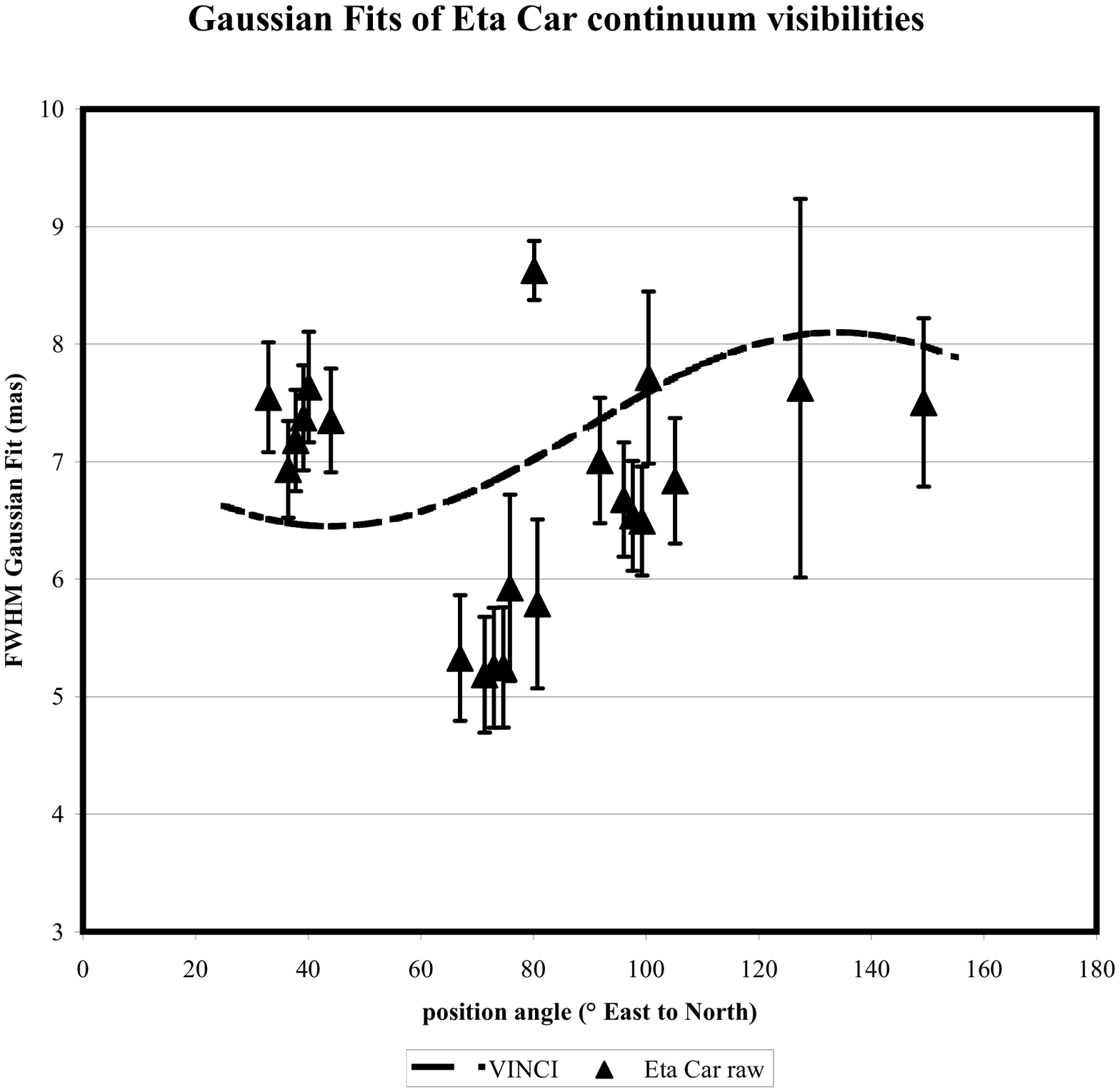} 
\includegraphics[width=5.7cm]{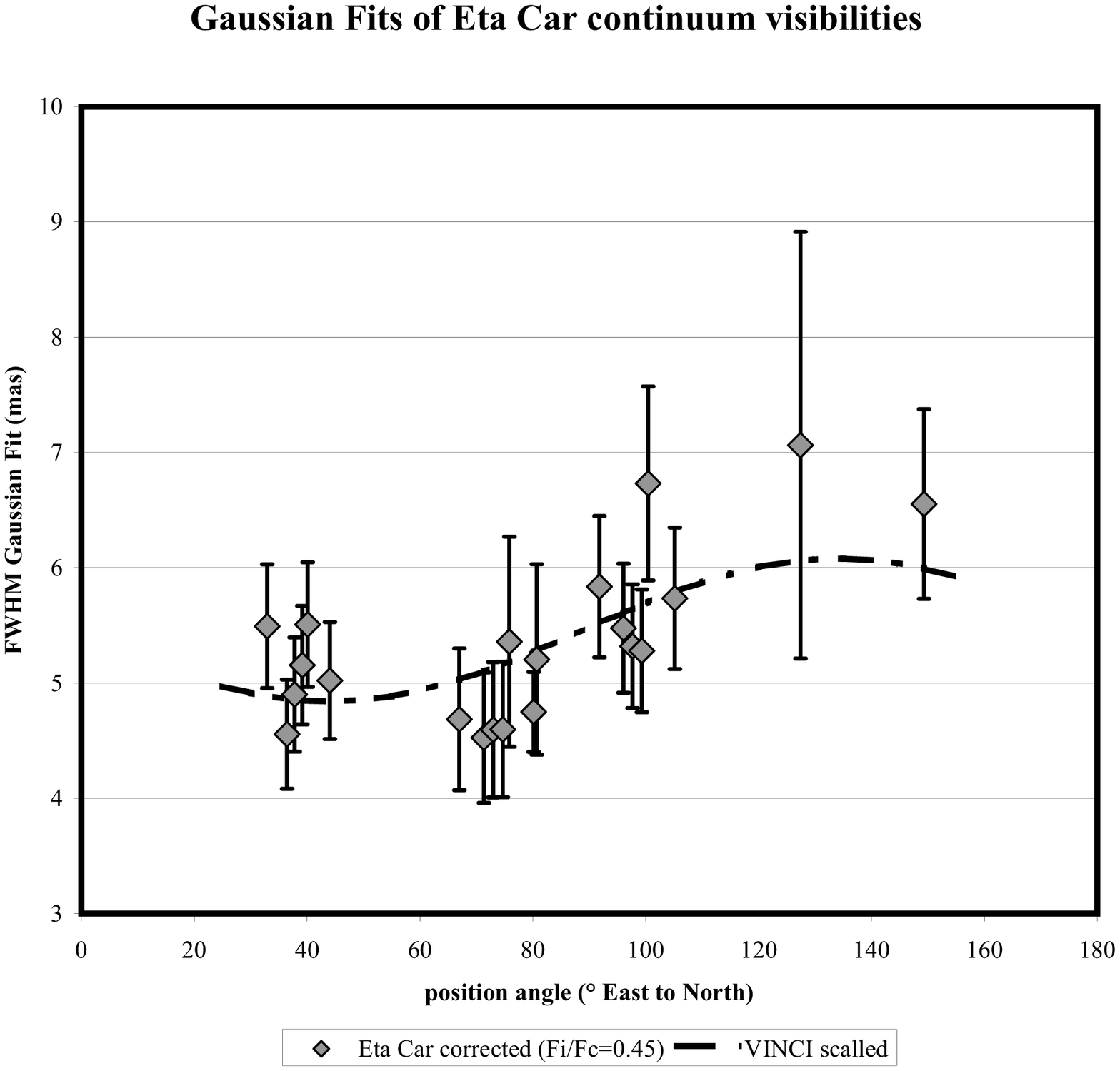} 
\caption{\scriptsize{
AMBER observations of $\eta Carinae$ in the continuum. FWHM of the Gaussian fit as a function of the position angle of the projected baseline. Left: from raw visibility (triangles) and comparison with the Van Boekel et al. fit of VINCI data (dashed curve). Right: from visibility corrected assuming that 31\% of the flux comes from an interferometrically fully resolved component (diamonds). The dashed curve shows the VINCI data fit scaled down by a factor 1.33 or assuming that in VINCI data the resolved component contributes to 47\% in the total flux instead of 43\% in the original VINCI fit.}}
\label{petrov:fig:2}
\end{figure}
The analysis of this continuum data confirms the importance and the difficulty of separating the contributions from different elements in the single mode field. A full solution would be to have interferometric observations with some field or at least to be able to have an efficient mosaicing strategy. Maybe the new infrared image sensor IRIS can be used for that purpose. The spectral information will help refining this task. We already know that the "resolved component" discussed here does not show sharp spectral features, unlike for example the Weigelt blobs which have been shown to generate narrow band features when we are pointing about 200 mas away from the brightest spot.
A more detailed discussion of the shape of the complex visibility through the line is quite premature and would make this paper even longer. A preliminary result is that the differential phase through the spectral lines is basically flat but for local spikes corresponding to photocenter displacements substantially smaller than the object size. This seems in contradiction with an optically thick wind with a surface on which the different equal velocity zones are clearly separated. In fact the data seem to show similar ovoid structures in all spectral channels, the key change being the optical depth.
\section{Conclusion}
\vskip -0.2cm
At an early stage in the reduction and interpretation of the first AMBER data on bright massive hot stars, we have tried to illustrate our strategy for model fitting in spite of an extremely limited u-v coverage. We want to use as much as possible the pre existing information and to find new features where our measures as a function of $\lambda$ differ significantly from the best model fitted in continuum data. For $\gamma^{2} Vel$ we are able to show that the interferometric signal is dominated by the binary system but that it is necessary to include an unresolved component with a spectra almost flat over the K band. However we also detect smaller scale structures in the system that are a good candidate for a signature of the wind-wind collision zone. The modeling of this zone will also allow to constrain further the nature of the spectral continuum component. For $\eta Car$, the situation is made more complex by the necessity to evaluate quite accurately the contribution of the larger scale structure to the flux collected by the fibers. The analysis of the AMBER data confirms the VINCI observation of a structure elongated in the direction of Homonculus nebulae, but it also shows that it is necessary to revise the VINCI evaluation of larger scale structure contribution. The consequences are quite important since a fairly limited variation ($<$5\%) of the contribution can change the estimated size by about 30\%. Next, we will try to combine AMBER measurements with spectrally resolved NACO+PF observations, to have a map as accurate as possible of the different scales of structures and will further analyze the differential and closure phases which are much less sensitive to large scale underlying structures
\vskip 0.1cm
\noindent
{\bf Acknowledgements: } The authors deeply acknowledge the AMBER consortium members, the staff of the associated Institutes\footnote{\noindent See list of consortium members and associates at: \\{\tt http://amber.obs.ujf-grenoble.fr/article.php3?id\_article=45}} and the ESO/VLTI team who permitted to obtain these results.
%
%

%
%

\end{document}